# Removing Leakage and Surface Recombination in Planar Perovskite Solar Cells

Kristofer Tvingstedt,*,† Lidón Gil-Escrig,‡ Cristina Momblona,‡ Philipp Rieder,† David Kiermasch,† Michele Sessolo,‡ Andreas Baumann,§ Henk J. Bolink,‡ and Vladimir Dyakonov†,§

†Experimental Physics VI, Julius Maximillian University of Würzburg, 97074 Würzburg, Germany
‡Instituto de Ciencia Molecular, Universidad de Valencia, C/Catedrático J. Beltrán 2, 46980 Paterna, Spain
§Bavarian Center for Applied Energy Research, 97074 Würzburg, Germany

Ⓢ Supporting Information

**ABSTRACT:** Thin-film solar cells suffer from various types of recombination, of which leakage current usually dominates at lower voltages. Herein, we demonstrate first a three-order reduction of the shunt loss mechanism in planar methylammonium lead iodide perovskite solar cells by replacing the commonly used hole transport layer poly(3,4-ethylenedioxythiophene):poly(styrenesulfonate) (PEDOT:PSS) with a better hole-selective polyarylamine. As a result, these cells exhibit superior operation under reduced light conditions, which we demonstrate for the extreme case of moonlight irradiance, at which open-circuit voltages of 530 mV can still be obtained. By the shunt removal we also observe the $V_{OC}$ to drop to zero after as long as 2 h after the light has been switched off. Second, at higher illumination intensities the dominant losses in the PEDOT:PSS-based cell are ascribed to surface recombination and are also proven to be substantially minimized by instead employing the polyarylamine. We attribute the reduced shunt and surface recombination to the far better suited semiconductor character of the polyarylamine, compared to that of PEDOT:PSS, efficiently blocking electrons from recombining at this electrode.

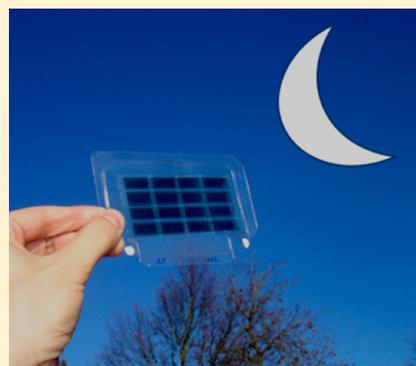

Solar cells are generally developed for operation under intense illumination such as direct sunlight or even using light concentrators. There are, however, numerous conditions in which the light intensity is lower because of nonideal orientation, sunrise or sunset, cloudy days, or indoor conditions. Importantly, to fully understand recombination mechanisms in solar cells, it is essential to also evaluate them under as large a range of illumination intensities as possible to assess which recombination pathway becomes dominant under each corresponding carrier concentration. Because recombination of excited charge carriers in photovoltaic devices are to various extents dependent on the charge carrier concentration, the rates accordingly depend on the illumination intensity itself. The general Shockley diode equation is of vital importance for solar cells because it defines how the total recombination processes depend on the internal voltage in a device. The relation between recombination current and voltage is usually described more accurately by the inclusion of an ideality factor ($n$), which varies depending on the dominant recombination mechanism. However, the Shockley diode equation is, even with the inclusion of an ideality factor, usually not sufficient to describe all processes that contribute to the overall recombination current dependency on the actual *external* voltage, $V$. Therefore, a more generalized Shockley equation (eq 1) accounting for both series and shunt resistances ($R_{Series}$ and $R_{Shunt}$, respectively) is required:

$$J = J_0(\exp[q(V - JR_{Series})/nkT] - 1) + \frac{V - JR_{Series}}{R_{Shunt}} - J_{Photo} \quad (1)$$

where $q$ is the electronic charge, $k$ the Boltzmann constant, and $J_0$ the dark saturation current. To identify the internal voltage drop over a diode it is important[1] to identify the influence of these resistive losses, which for high voltages are governed by the series resistance[2] and for low voltages by the shunt resistance.[3] The loss mechanism over shunt paths is generally referred to as "leakage current" because it profoundly increases the recombination current to values considerably larger than expected from the ideal diode. However, these losses are not associated with the material itself, but rather with the entire device configuration. Unfortunately, most thin-film solar cells today suffer from such leakage losses.[3−5] For cells operating under stronger lighting conditions, such as solar intensities, other recombination loss mechanisms become dominant, and shunt paths do not usually have a noticeable effect on the device performance, except if the shunt loss is extraordinary high. For cells that operate under weaker illumination conditions,[6−8] or when used as sensitive photodetectors, it is however decisive that the recombination over the shunt is very low. More important from a scientific perspective is that the presence of shunt resistance frequently hinders the correct assignment of material representative carrier recombination dynamics, for example, as determined from electrical transient behavior of solar cells.[9] It is important to note that the dependency of the open-circuit voltage ($V_{OC}$) on light intensity ($P_L$) is also heavily affected by shunt recombination losses.[1] Such mechanisms have in fact earlier very often been misinterpreted and led to erroneous assignments of unrealistically high ideality factors from the $V_{OC}(P_L)$ relation. Recent studies have shown that similar shunt formation statistics are found for rather different photovoltaic technologies.[4]

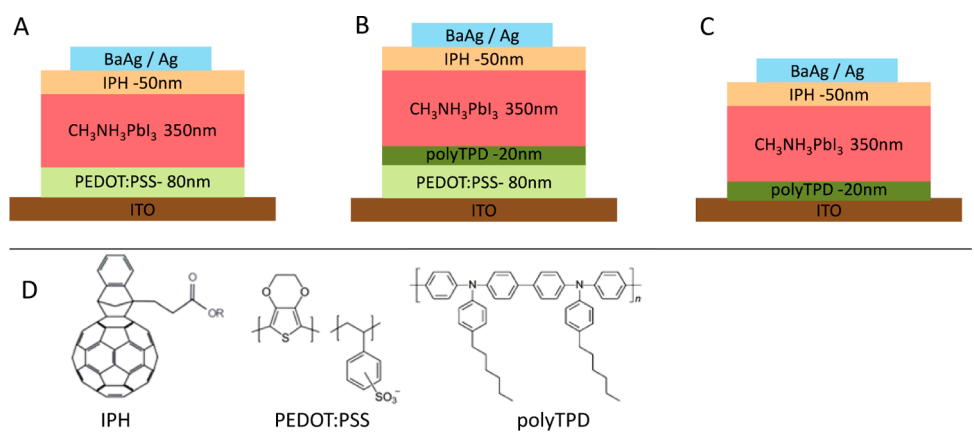

Figure 1. Device layouts of the studied planar methylammonium lead iodide cells employing three different hole transport layers: device A, 80 nm of PEDOT:PSS; device B, with 80 nm of PEDOT:PSS and 20 nm of polyTPD; and device C, with 20 nm of polyTPD only. The perovskite layers and top electron-selective layers are identical as they were prepared in the same run. The chemical structure of the charge transport materials is outlined in panel D.

Lead halide perovskites have attracted widespread attention because of their outstanding suitability as photovoltaic energy converters[10−13] and photodetectors[14−17] with power conversion efficiencies exceeding 22%. Currently, perovskite solar cells are manufactured in two main configuration categories, mesoporous or planar.[10] Whereas mesoporous cells employ scaffolds of (usually $TiO_x$) nanoparticles as electron selective layer, the choice of selective layers for the planar configuration is more open. Poly[N,N′-bis(4-butylphenyl)-N,N′-bis(phenyl)-benzidine (polyTPD) is a poly(triaryl)amine, which because of its transport property and thermal stability has been used as a hole-selective material in mainly organic light-emitting diodes[18] and lately in photovoltaics. While the benefits of this material have been demonstrated by Malinkiewicz et al.[19] and Zhao et al.[20] for planar perovskite cells at solar illumination intensities, the reason for the better performance has not been clarified, and the connection to the herein observed reduced leakage current has not yet been widely noted. However, Fang and Huang did recently observe a reduction in the leakage current of perovskite photodetectors with a cross-linkable arylamine as hole transporter.[14] The earlier work clearly proved that polyTPD on top of poly(3,4-ethylenedioxythiophene):poly-(styrenesulfonate) (PEDOT:PSS) substantially improves the performance at sun intensities, and the differences were mostly ascribed to crystal grain size alterations affecting trap concentration.[20] Recent work by El Labban et al.[21] and Hou et al.[22] proposed instead that electrons in the perovskite are parasitically lost at the PEDOT:PSS interface. Herein, we simply remove one layer in a well-known[19] planar methylammonium lead iodide perovskite stack and demonstrate first an unprecedented reduction of the parasitic shunt losses, rendering the perovskite cell suitable for operation also under extremely low irradiance conditions. Second, we confirm that the dominant loss mechanism in the PEDOT:PSS-based solar cells can be ascribed to surface recombination, preventing the voltage from increasing with increased illumination intensities, as opposed to the superior polyTPD-based devices. Finally, we propose an explanation for why the latter devices do not suffer from such shortcomings.

We manufactured 16 planar perovskite solar cells for each of the three device configurations presented in Figure 1, according to methods detailed in the Supporting Information. The first set of devices (type A) comprised a PEDOT:PSS film spin-coated on top of indium tin oxide (ITO). Our second set (type B) comprised a thin film of polyTPD deposited on top of the PEDOT:PSS layer, whereas our third set (type C) used a polyTPD film alone, deposited directly on ITO. The representative devices reported in the Letter, with active area of 6.5 mm², were manufactured in one run in the evaporation chamber, rendering identical uniform perovskite films as confirmed via scanning electron and optical microscopy (Figures S1 and S2). Figure 2 displays the dark J–V measurements of the cells and of a reference Si photodetector, S1787-04, from Hamamatsu. This Si photodetector,[23] with a shunt resistance of typically 100 GΩ, is highly suited for photodetection over several orders of light intensities. The general influence of shunt resistances on dark diode J–V curves is also demonstrated by placing a 1 MΩ resistor in parallel with the Si photodetector, essentially removing the five lowest orders in photosensitivity.

The PEDOT:PSS/polyTPD-based cells (device B) shows a very commonly observed shunt resistance of ∼70 MΩ providing a leakage current of around 0.45 mA m$^{-2}$ at 0.2 V. A 70 MΩ resistor well describes the entire voltage range of the losses due to the shunt path. Such values are also representative for organic solar cells likewise employing PEDOT:PSS.[1] The perovskite solar cells with a bare PEDOT:PSS layer (device A) displayed a slightly lower shunt resistance of ∼15 MΩ. Device C (with only polyTPD) shows on the other hand a remarkably reduced leakage current and a corresponding shunt resistance as high as 15 GΩ, comparable with the Si photodetector. However, the assumption of the shunt being solely described by an ohmic resistor might not fully apply for device C, as we note minor deviations from linear behavior in the forward sweep. While it is indeed possible to better fit a larger voltage range with a proposed[4,5] power law expression, we prefer here to stay with the linear approximation to enhance clarity. We note that the effective leakage current of this device amounts to merely 2.1 $\mu$A m$^{-2}$ at ±0.2 V. It is important to point out that the shunt value extracted from the polyTPD-based cells was very reproducible, with not one of the 16 manufactured devices showing deviations larger than 16% from the 15 GΩ resistance presented in Figure 2.

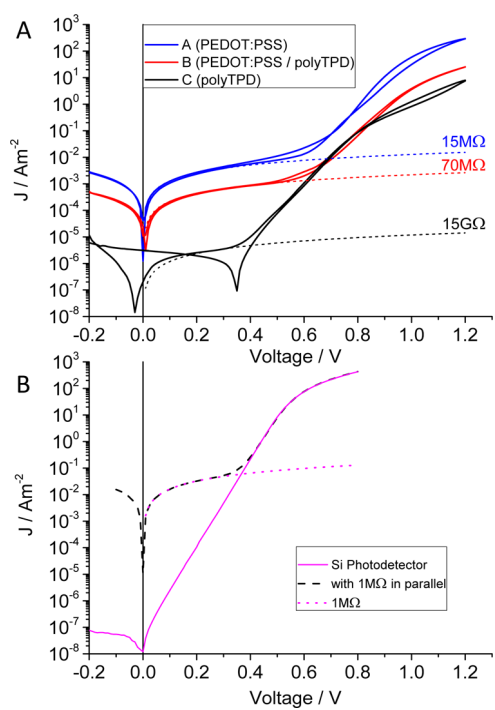

Figure 2. (A) Dark $J$–$V$ curves of the three studied photovoltaic diodes. The device employing only polyTPD (device C) is characterized by a leakage current 3 orders of magnitude lower than device A employing only PEDOT:PSS. (B) A reference 6.2 mm$^2$ Hamamatsu photodetector with nominal ~100 GΩ built-in shunt resistor was deliberately shunted by placing a 1 MΩ resistor in parallel with the device.

Evaluating the double-sweep $J$–$V$ curves of Figure 2, we note the almost complete lack of hysteresis in the exponential part of the curve for the polyTPD-based cells (devices B and C). Current voltage hysteresis is otherwise an unfortunate and common feature[24] of perovskite solar cells, and its exact origin remains to be determined. The graphs presented here are logarithmic $J$–$V$ curves, substantially raising the bar for claiming the property of being "hysteresis-free". In the ultralow current regime of device C, with only polyTPD, some form of hysteresis can however be clearly identified. We attribute this difference to charging of the cells during or after the reverse sweep. As a result of this, an "open-circuit voltage in the dark" can be detected, albeit only in the reverse sweep. This at a first glance highly anomalous solar cell behavior may however not be so unexpected considering that the fast leakage path is so effectively removed, and it is plausible that some charges, stored somewhere in the device, have simply not recombined in the studied time frame. We therefore chose to conduct a study of the time decay of the open-circuit voltage (OCVD). As noted earlier,[9,25] it is then crucial to employ as high as possible input impedance to correctly measure the solar cell, as opposed to unintentionally measuring the voltage decay over the utilized measurement instrumentation itself (see the Supporting Information for further details). We illuminate the three perovskite solar cells and the Si photodiode with light intensities corresponding to room light for 5 s, because this was sufficient to ensure a steady-state starting condition where the generation rate equals the recombination rate, and then switch off the light ($t = 10^{-6}$ s) and monitor the decay as presented with a lin.–log scale in Figure 3.

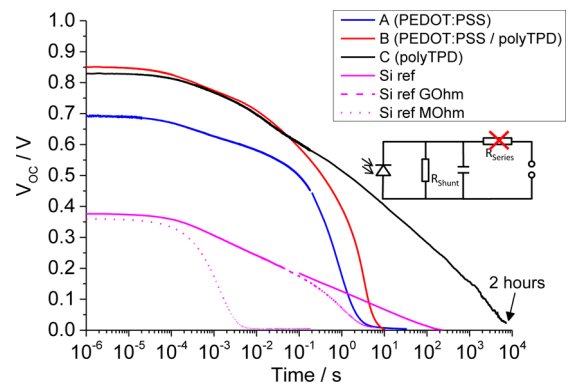

Figure 3. Open-circuit voltage decay for the three perovskite cells and the Si reference photodetector, the latter with different measurement resistances in parallel. The shunt-free perovskite device C displays charge carriers remaining at the electrodes of the device even after up to as much as 2 h. Inset shows the equivalent circuit used in the corresponding fitting (Figure S3).

As comparison, we also include the OCVD traces for the Si photodiode obtained when employing either the built-in megaohm of the oscilloscope or the gigaohm load for the longer time scales and lower voltages. The rapid voltage drops observed in those cases are occurring over the employed measurement instrumentation and not inside the Si diode itself. All perovskite cells are instead evaluated with a high load resistance, and the voltage drop occurring after around 1 s is therefore due to the shunt resistance of the cell itself. Device C, using only polyTPD as the hole-selective layer, does not display such a drop at all but decays instead over at least three more decades in time compared to the PEDOT:PSS-based cells. Device C retains a $V_{OC}$ of 0.25 V even 3 min after the light has been switched off and does not reach 0 V until after 2 h. The large differences in time it takes for the voltage to reach zero clearly highlight the influence that the often observed "leakage current" generally has on the behavior of solar cells. The equivalent circuit of the solar cell is described in the inset of Figure 3 and includes a diode, capacitor, and shunt resistor in parallel. As we study only open-circuit conditions, the value of the also present series resistance becomes unimportant here. With this circuit in mind, we can assign the long and rather uniform decay of voltage with the logarithm of time as a capacitive contribution from charges probably stored on some of the electrode layers of the device. These charge carriers, contributing to the overall voltage, do not disappear as fast as carriers in the bulk of the device because they are separated from each other by the perovskite layer. They instead represent capacitive charges that can decay via either the parasitic pathway that the shunt represents, that is, over an RC component with a traditional monoexponential decay characteristic, or via the wide bandgap semiconductor diode that the bulk of the perovskite itself represents. Depending on the value of the shunt resistance, the full voltage decay of Figure 3 can be quite accurately described by one of the two routes obtained by simply equating the time derivative of charges on the capacitor with either the current through a resistor or through a diode. Equation 2 represents the well-known RC identity and its solution, whereas eq 3 represents instead the diode−capacitor[26] identity and its solution:

$$-C\frac{dV}{dt} = \frac{V}{R_{Shunt}} \text{ with solution } V(t) = V_{t=0}\,e^{-t/RC} \quad (2)$$

$$-C\frac{dV}{dt} = I_0(e^{V/nkT} - 1) \text{ with solution}$$
$$V(t) = -nkT\ln(1 - (1 - e^{-V_{t=0}/nkT})e^{-(I_0/nkT\cdot C)t}) \quad (3)$$

where $V_{t=0}$ is the starting steady-state voltage value; $n$ and $I_0$ are the ideality factor and dark saturation current of the diode, respectively. We emphasize here that this diode−capacitor decay is very different from a RC decay, because it follows a linear behavior with the logarithm of time for the larger part of the time axis. Accordingly, we have chosen to plot the OCVD decay in Figure 3 on a linear voltage axis and a logarithmic time axis. Fits of the measured OCVD traces to eqs 2 and 3 are included in the Supporting Information (Figure S3) where corresponding capacitance values are also provided. The end time of such diode−capacitor decays, that is, when the voltage reaches zero and the capacitor is discharged, is from eq 3 ruled by the simple ratio of the capacitance and the dark saturation current according to[26]

$$t_{End} = \frac{nkTC}{I_0} \quad (4)$$

Very long decay end times can therefore be reached only by shunt-free devices such as device C and the Si reference cell, whereas the other perovskite devices enter a faster RC regime at much earlier times, when the shunt starts to rule the dynamics. Ions have earlier been assigned to be responsible for the hysteresis, but recently also to anomalously long-lived charges.[27] Ionic displacement would possibly be more consistent with the voltage persistency also recently observed[9] but not seen in the low-intensity decay trace of Figure 3. Acknowledging the possibility of ionic displacement affecting the voltage decay, we would however expect more similar behavior in the OCVD decay for all three device structures, because they comprise identical perovskite layers. The Si cell also requires 115 s to reach zero volts, which obviously should not be associated with any ionic effect. We therefore note that long decays do not necessarily have to do with the presence of ions, although the magnitude of the capacitance may indeed be partly related to their presence.

Having confirmed that the voltage decay is indeed extremely slow at lower voltages, when the leakage recombination path has been so effectively removed, we can justify the observed "$V_{OC}$ in the dark", in the low-voltage reverse sweep of Figure 2A, considering that the J−V measurement is just sweeping the voltage faster here than the capacitive decay rate itself. Hence, even if our sweep rate is as low as 9 mV s$^{-1}$, a substantial amount of the charges at 0.4 V is still remaining at the electrodes of the device when we, in our reverse sweep 11 s later, are measuring 0.3 V. Figure S6 reveals further the impact of different sweep end voltage values and sweep rates for the polyTPD-based cell.

The labeling "leakage current" can also be somewhat misleading as it may be interpreted as an extra current flow originating purely from the applied voltage of an external electrical power source. To emphasize that leakage current over shunt paths, irrespective of their origin, indeed represents a true low-light intensity recombination mechanism, the *steady-state* open-circuit voltage under a large set of illumination intensities is presented in Figure 4A. The recombination over the shunt path can here clearly be identified as the rapid voltage drop at lower light intensity conditions (for the PEDOT:PSS-based cells around 10$^{-4}$ suns). In these devices, the quasi-Fermi levels cannot split properly according to what is expected from the Shockley equation, because an additional faster loss mechanism prevents a sufficient carrier density to be sustained. However, this is clearly not the case for the polyTPD-based cell, which shows no voltage drop even at the lowest intensities available in our setup (~10$^{-7}$ suns). The whole relationship between $V_{OC}$ and intensity is instead well-explained without the extra shunt term, such that the position of the quasi-Fermi levels can be fully assigned to bulk carrier recombination even at low voltages for the polyTPD-based cell. The overall similarity among Figures 2−4 in fact elucidates that the steady-state (Figure 4), quasi steady-state (Figure 2), and the full voltage decay measurements (Figure 3) indeed are probing identical features.

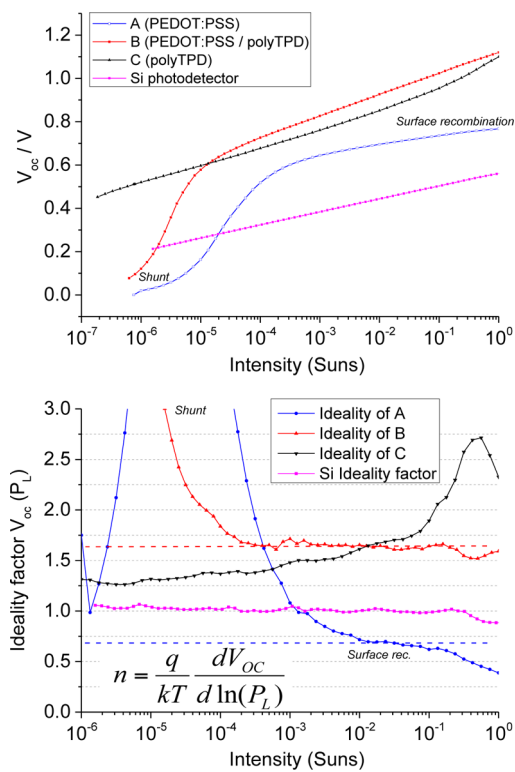

Figure 4. (A) Steady-state $V_{OC}$ vs illumination intensity ($P_L$, expressed in sun equivalent units). Devices A and B, containing PEDOT:PSS, display a rapid voltage drop at low light intensity due to the presence of the additional shunt recombination pathway. (B) Ideality factors as determined from the inset equation and the relationship in panel A clearly revealing the different loss mechanisms ruling the devices at higher light intensities.

For the higher-intensity regime in Figure 4A, we clearly also note a large difference in the overall slope of the $V_{OC}(P_L)$ relation for device A and accordingly that this device is not able to reach the $V_{OC}$ values of the polyTPD-based cells at any light intensity. The slope represents the series resistance-free ideality factor and is a strong indicator of what type of recombination is ruling the cells.[1] The ideality factor plotted as a function of light intensity in Figure 4B reveals exceptionally low values for the PEDOT:PSS-based cell. Similar notes of low ideality factors for such cells were also made recently by Zhao et al.,[20] but the reason behind it was not assigned. Low values of ideality factors in conjunction with overall very high values of dark saturation currents are actually a strong indicator of a pronounced surface recombination.[1] The polyTPD-based cells are not limited by this but are instead, as generally often stated for higher efficiency perovskites, limited by trap assisted recombination. Strong surface recombination in PEDOT:PSS-based perovskites was also very recently proposed by Hou et al.,[22] whereas the origin of the surface recombination was not proposed. We herein assign the reduced surface recombination and lower leakage of device C primarily to the superior semiconducting character of polyTPD with respect to that of PEDOT:PSS. Electrons present in the perovskite conduction band cannot make the transition into polyTPD, because of its very high lowest unoccupied molecular orbital level of approximately −2.4 eV.[19] On the other hand, PEDOT:PSS displays instead characteristics analogous to semimetals[28] and does not qualify as an efficient electron-blocking material. To distinguish the clear difference in energetic distribution of states in the two studied materials, Figure 5 displays their absorption coefficient on a logarithmic scale together with the external quantum efficiency (EQE) of a perovskite (device C) cell.

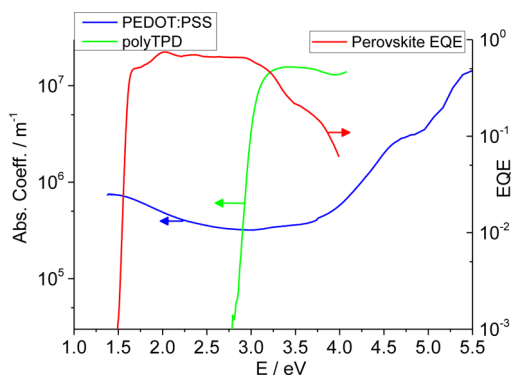

Figure 5. Absorption coefficient of the two studied hole-transport materials highlighting that only polyTPD has a real band edge. The EQE of device C is included as a textbook example of a clear band edge semiconductor material.

The clear presence of low energetic transitions allowed in PEDOT:PSS cannot be identified in polyTPD. This shortcoming of PEDOT:PSS allows a non-negligible part of the perovskite electrons to be parasitically extracted via midgap states,[29] such as polaron or bipolaron states, very abundant therein.[28] A similar proposal was made recently when replacing PEDOT:PSS with CuSCN in organic solar cells.[30] In contrast, the very steep absorption edge of polyTPD indicates instead that no such gap states mediating electrode recombination are present in this material. We therefore herein propose that not only the main absorbing material should possess as steep a band-edge[31] as possible, but necessarily also the employed charge-selective electrodes, in order to ensure that the density of midgap states is minimized also in these layers. PolyTPD is an example of one material that features these properties.

We note also that device B, with the PEDOT:PSS underneath the polyTPD, apparently partly removes the excellent low leakage current of device C. Optical microscopy (see the Supporting Information) reveals micrometer-sized spots located underneath the uniform perovskite film of device B. We attribute these to limited wetting of polyTPD on PEDOT:PSS, and we propose these sites, not visible in scanning electron microscopy images (see the Supporting Information) of the perovskite surface itself, also act as possible shunt formations. Similar features were recently[32] observed using hyperspectral imaging and indeed were associated with a reduced local quasi Fermi-level splitting. These features are not present in devices A and C. The Supporting Information further displays J–V curves under a set of selected familiar light intensities for the three evaluated devices. We show for example that only the shunt-free device C is able to retain an unprecedented open-circuit voltage of **530 mV** under moonlight intensities.

In conclusion, our message with this contribution is threefold: We first suggest to exclude PEDOT:PSS as a hole transport material because of its semimetallic nature and corresponding abundancy of intergap states that increases recombination at this interface and further leads to large leakage currents. We conclude that this loss mechanism prevents the difference of the two perovskite quasi-Fermi levels from maintaining a high value at both high and low irradiance levels, but for rather different reasons. At low voltage values, the shunt loss rules, while at higher values, a pronounced contribution of surface recombination instead becomes dominant. PolyTPD does not show any of these features but acts instead as an almost perfect hole-selective layer that excellently inhibits electron recombination at this side of the device, independent of voltage. Second, from our OCVD measurements, we conclude that the shunt-free devices show a capacitive contribution to voltage that remains for up to 2 h after the light has been switched off and that can be fully described with a capacitor discharging over a diode. Third, we propose the steepness of band edges as determined from sensitive absorption profiles as a figure of merit for a good charge-selective layer.

## ■ ASSOCIATED CONTENT

### ⓈSupporting Information

The Supporting Information is available free of charge on the ACS Publications website at DOI: 10.1021/acsenergylett.6b00719.

> Experimental details of the manufacturing procedures, device statistics, and measurement methodology details; SEM and optical images of the manufactured devices taken after the top electrode has been delaminated; equivalent circuit fitting of the OCVD; J–V sweep alterations; J–V characteristics of the four devices evaluated under a set of selected familiar light intensities. (PDF)

## AUTHOR INFORMATION

**Corresponding Author**
*E-mail: ktvingstedt@physik.uni-wuerzburg.de.

**ORCID**
Kristofer Tvingstedt: 0000-0003-0516-9326
Vladimir Dyakonov: 0000-0001-8725-9573

**Notes**
The authors declare no competing financial interest.

## ACKNOWLEDGMENTS

K.T., D.K., P.R., A.B., and V.D. acknowledge the BMBF for funding through the grant HYPER with Grant Agreement Number 03SF0514. L.G.-E., C.M., M.S., and H.J.B acknowledge financial support from the Spanish Ministry of Economy and Competitiveness (MINECO) via the Unidad de Excelencia María de Maeztu MDM-2015-0538, MAT2014-55200, and PCIN-2015-255 and the Generalitat Valenciana (Prometeo/2012/053). C.M. and M.S. thank the MINECO for pre- and postdoctoral (JdC) contract. The authors also acknowledge constructive criticism from the reviewers on the first version of the manuscript.

## REFERENCES

(1) Tvingstedt, K.; Deibel, C. Temperature Dependence of Ideality Factors in Organic Solar Cells and the Relation to Radiative Efficiency. *Adv. Ener. Mater.* **2016**, *6*, 1502230.

(2) Fong, K. C.; McIntosh, K. R.; Blakers, A. W. Accurate Series Resistance Measurement of Solar Cells. *Prog. Photovoltaics* **2013**, *21*, 490−499.

(3) Dongaonkar, S.; Servaites, J. D.; Ford, G. M.; Loser, S.; Moore, J.; Gelfand, R. M.; Mohseni, H.; Hillhouse, H. W.; Agrawal, R.; Ratner, M. A.; Marks, T. J.; Lundstrom, M. S.; Alam, M. A. Universality of Non-Ohmic Shunt Leakage in Thin-Film Solar Cells. *J. Appl. Phys.* **2010**, *108*, 124509.

(4) Dongaonkar, S.; Loser, S.; Sheets, E. J.; Zaunbrecher, K.; Agrawal, R.; Marks, T. J.; Alam, M. A. Universal Statistics of Parasitic Shunt Formation in Solar Cells, and Its Implications for Cell to Module Efficiency Gap. *Energy Environ. Sci.* **2013**, *6*, 782−787.

(5) Liao, Y. K.; Kuo, S. Y.; Hsieh, M. Y.; Lai, F. I.; Kao, M. H.; Cheng, S. J.; Chiou, D. W.; Hsieh, T. P.; Kuo, H. C. A Look into The Origin of Shunt Leakage Current of Cu(In,Ga)Se-2 Solar Cells via Experimental and Simulation Methods. *Sol. Energy Mater. Sol. Cells* **2013**, *117*, 145−151.

(6) Steim, R.; Ameri, T.; Schilinsky, P.; Waldauf, C.; Dennler, G.; Scharber, M.; Brabec, C. J. Organic Photovoltaics for Low Light Applications. *Sol. Energy Mater. Sol. Cells* **2011**, *95*, 3256−3261.

(7) Kawata, K.; Tamaki, K.; Kawaraya, M. Dye-Sensitised and Perovskite Solar Cells as Indoor Energy Harvestors. *J. Photopolym. Sci. Technol.* **2015**, *28*, 415−417.

(8) Mathews, I.; King, P. J.; Stafford, F.; Frizzell, R. Performance of III-V Solar Cells as Indoor Light Energy Harvesters. *Ieee J. Photovolt.* **2016**, *6*, 230−235.

(9) Baumann, A.; Tvingstedt, K.; Heiber, M. C.; Vath, S.; Momblona, C.; Bolink, H. J.; Dyakonov, V. Persistent Photovoltage in Methylammonium Lead Iodide Perovskite Solar Cells. *APL Mater.* **2014**, *2*, 081501.

(10) Grätzel, M. The Light and Shade of Perovskite Solar Cells. *Nat. Mater.* **2014**, *13*, 838−842.

(11) Lee, M. M.; Teuscher, J.; Miyasaka, T.; Murakami, T. N.; Snaith, H. J. Efficient Hybrid Solar Cells Based on Meso-Superstructured Organometal Halide Perovskites. *Science* **2012**, *338*, 643−647.

(12) Jeon, N. J.; Noh, J. H.; Yang, W. S.; Kim, Y. C.; Ryu, S.; Seo, J.; Seok, S. I. Compositional Engineering of Perovskite Materials for High-Performance Solar Cells. *Nature* **2015**, *517*, 476−480.

(13) Zuo, C.; Bolink, H. J.; Han, H.; Huang, J.; Cahen, D.; Ding, L. Advances in Perovskite Solar Cells. *Adv. Sci.* **2016**, *3*, 1500324.

(14) Fang, Y.; Huang, J. Resolving Weak Light of Sub-Picowatt per Square Centimeter by Hybrid Perovskite Photodetectors Enabled by Noise Reduction. *Adv. Mater.* **2015**, *27*, 2804−2810.

(15) Saidaminov, M. I.; Adinolfi, V.; Comin, R.; Abdelhady, A. L.; Peng, W.; Dursun, I.; Yuan, M.; Hoogland, S.; Sargent, E. H.; Bakr, O. M. Planar-Integrated Single-Crystalline Perovskite Photodetectors. *Nat. Commun.* **2015**, *6*, 8724.

(16) Fang, Y.; Dong, Q.; Shao, Y.; Yuan, Y.; Huang, J. Highly Narrowband Perovskite Single-Crystal Photodetectors Enabled by Surface-Charge Recombination. *Nat. Photonics* **2015**, *9*, 679−686.

(17) Dong, R.; Fang, Y.; Chae, J.; Dai, J.; Xiao, Z.; Dong, Q.; Yuan, Y.; Centrone, A.; Zeng, X. C.; Huang, J. High-Gain and Low-Driving-Voltage Photodetectors Based on Organolead Triiodide Perovskites. *Adv. Mater.* **2015**, *27*, 1912−1918.

(18) Sun, Q. J.; Hou, J. H.; Yang, C. H.; Li, Y. F.; Yang, Y. Enhanced Performance of White Polymer Light-Emitting Diodes Using Polymer Blends as Hole-Transporting Layers. *Appl. Phys. Lett.* **2006**, *89*, 153501.

(19) Malinkiewicz, O.; Yella, A.; Lee, Y. H.; Espallargas, G. M.; Graetzel, M.; Nazeeruddin, M. K.; Bolink, H. J. Perovskite Solar Cells Employing Organic Charge-Transport Layers. *Nat. Photonics* **2013**, *8*, 128−132.

(20) Zhao, D.; Sexton, M.; Park, H.-Y.; Liu, S.; Baure, G.; Nino, J. C.; So, F. High-Efficiency Solution-Processed Planar Perovskite Solar Cells with a Polymer Hole Transport Layer. *Adv. Ener. Mater.* **2015**, *5*, 1401855.

(21) El Labban, A.; Chen, H.; Kirkus, M.; Barbe, J.; Del Gobbo, S.; Neophytou, M.; McCulloch, I.; Eid, J. Improved Efficiency in Inverted Perovskite Solar Cells Employing a Novel Diarylamino-Substituted Molecule as PEDOT:PSS Replacement. *Adv. Ener. Mater.* **2016**, *6*, 1502101.

(22) Hou, Y.; Chen, W.; Baran, D.; Stubhan, T.; Luechinger, N. A.; Hartmeier, B.; Richter, M.; Min, J.; Chen, S.; Quiroz, C. O. M.; et al. Overcoming the Interface Losses in Planar Heterojunction Perovskite-Based Solar Cells. *Adv. Mater.* **2016**, *28*, 5112.

(23) SI Photodiode S1787-04. http://www.hamamatsu.com/eu/en/product/alpha/P/4103/S1787-04/index.html.

(24) Snaith, H. J.; Abate, A.; Ball, J. M.; Eperon, G. E.; Leijtens, T.; Noel, N. K.; Stranks, S. D.; Wang, J. T.-W.; Wojciechowski, K.; Zhang, W. Anomalous Hysteresis in Perovskite Solar Cells. *J. Phys. Chem. Lett.* **2014**, *5*, 1511−1515.

(25) Sudheendra Rao, K.; Mohapatra, Y. N. Open Circuit Voltage Decay Transients and Recombination in Bulk-Heterojunction Solar Cells. *Appl. Phys. Lett.* **2014**, *104*, 203303.

(26) Hellen, E. H. Verifying the Diode-Capacitor Voltage Decay. *Am. J. Phys.* **2003**, *71*, 797.

(27) O'Regan, B. C.; Barnes, P. R. F.; Li, X. E.; Law, C.; Palomares, E.; Marin-Beloqui, J. M. Optoelectronic Studies of Methylammonium Lead Iodide Perovskite Solar Cells with Mesoporous $TiO_2$: Separation of Electronic and Chemical Charge Storage, Understanding Two Recombination Lifetimes, and the Evolution of Band Offsets during J-V Hysteresis. *J. Am. Chem. Soc.* **2015**, *137*, 5087−5099.

(28) Bubnova, O.; Khan, Z. U.; Wang, H.; Braun, S.; Evans, D. R.; Fabretto, M.; Hojati-Talemi, P.; Dagnelund, D.; Arlin, J. B.; Geerts, Y. H.; et al. Semi-Metallic Polymers. *Nat. Mater.* **2014**, *13*, 190−194.

(29) Ratcliff, E. L.; Zacher, B.; Armstrong, N. R. Selective Interlayers and Contacts in Organic Photovoltaic Cells. *J. Phys. Chem. Lett.* **2011**, *2*, 1337.

(30) Treat, N. D.; Yaacobi-Gross, N.; Faber, H.; Perumal, A. K.; Bradley, D. D. C.; Stingelin, N.; Anthopoulos, T. D. Copper thiocyanate: An Attractive Hole Transport/Extraction Layer for Use in Organic Photovoltaic Cells. *Appl. Phys. Lett.* **2015**, *107*, 013301.

(31) Tvingstedt, K.; Malinkiewicz, O.; Baumann, A.; Deibel, C.; Snaith, H. J.; Dyakonov, V.; Bolink, H. J. Radiative Efficiency of Lead Iodide Based Perovskite Solar Cells. *Sci. Rep.* **2014**, *4*, 6071.

(32) El-Hajje, G.; Momblona, C.; Gil-Escrig, L.; Ávila, J.; Guillemot, T.; Guillemoles, J.-F.; Sessolo, M.; Bolink, H. J.; Lombez, L. Quantification of Spatial Inhomogeneity in Perovskite Solar Cells by Hyperspectral Luminescence Imaging. *Energy Environ. Sci.* **2016**, *9*, 2286−2294.